# Information Diffusion issues

By Jonathan Helmigh <johelmgh@gmail.com>

## [1]. INTRODUCTION

In this report there will be a discussion for Information Diffusion. There will be discussions on what information diffusion is, its key characteristics and on several other aspects of these kinds of networks.

This report will focus on peer to peer models in information diffusion. There will be discussions on epidemic model, OSN and other details related to information diffusion.

## [2]. INFORMATION DIFFUSION

Information diffusion refers to the process of spreading information. Now information spreads through networks and systems. A diffusion network is a network through which information is diffused. There are several factors that influences information diffusion. Primarily these factors are, strength of the connections among the nodes of the diffusion network and the structure, topology of the network. There are some metrics that are used to measure the quality of information diffusion. Network structure is evaluated based on the density, clustering, community structure, connected components, degree of distribution, weighted strength of the connections, strength of influence, communication frequency, spreading agent, type of information and attractiveness. A strong 'tie'

or connection should have affinity, more mutual and frequent contracts. [1]

Diffusion of information can range from simple to complex forms. In the simple form, it is between person to person and in complex form, individuals with thresholds are engaged. Social networks play significant role in information diffusion. Innovation is another outcome of information diffusion. But sometimes it is hindered. The 'influence' is spread through social media and its members in the form of information, opinion, innovation, ideas etc. In businesses, this is very important. They utilize this concepts in marketing and earning more revenue.

## [3]. BACKGROUND

Ad hoc and peer to peer networks, both share some common pitfalls. Problems in delivering messages in a network that has a rapidly changing topology. There are two methods of message delivery approaches. These two are opposite in nature.

In the first approach, a node in the network communicates with others proactively by active data transmission and inquiry. On the other approach, the communication link is created on demand.

In mobile peer to peer networks both approaches are applicable however, there will be significant overhead with proactive approaches when compared to the payload overhead in case of a dynamic network structure. The dynamic changes in the mobility is very common and basic in case of mobile networks. The nodes in mobile networks changes very frequently and the topology also changes. It needs to update the nodes between source and destination very frequently. Information will be available to all nodes in the network irrespective of the fact that whether the nodes are being beneficial from the information or not.

In case of common IP networks, a message is transmitted through the nodes in the network. There is no direct link between nodes. But in the point to point mobile networks, as the topology changes every now and then, so there are movements of the nodes. So, the system needs only networks with direct and single hop links. Information can be transmitted from source to destination through intermediate nodes. Network assumes the movement of the mobile nodes. Because of simplicity, in rest of the report, single hop communication links will be considered. [2]

# [4]. PEER TO PEER MODELS

In case of distribution of digital media over peer to peer networks, the users get information from others who has the information. Thus the process is decentralized and distributed in the nature. There are several features of this kind of sharing and information diffusion. Now, in a typical case, only a fraction of users have the content. There are many other users who seeks and receives the content from the holding users. Now, most of the users don't share it after getting it. So, the demand for the content is never full filled.

There are approaches in designing a peer to peer model for social networks that helps in overcoming such problems. This is mixed approach. The P2P network is used to reflect the distribution of the content from recent and current adopters, fulfillment of the demand etc. The model uses a payment model, where the distributors of the content gets paid. [3]

# [5]. PEER TO PEER EPIDEMIC MODEL

Information diffusion in distributed systems is implemented using epidemic algorithms. These algorithms are involved in propagating pairwise updates through networks. Epidemic algorithms are fully randomized and distributed in nature. During a session of information diffusion a peer takes a subset of rest of the peers randomly. Then propagates updates to those peers in several periodic rounds. The benefits of this 'epidemic' approach are robustness, scalability and consistency. There are several epidemic algorithms in case of information diffusion.

An early example of information diffusion and epidemic algorithm has been found in the process of propagating updates in database replication. Later on several modifications of the algorithm has been found in direct mailing, replication of servers, detection of network failures, garbage collection multicasting, distributed information management, buffer management, network awareness, message filtering etc. [4]

## 5.1 ANTI-ENTROPY

Anti-entropy is an epidemic process where information diffusion happens periodically using fixed time periods. The time frames having the fixed time periods are usually larger than the maximum round trip time between the participating peers. These are called rounds.

During each round, a peer takes another peer randomly. Then exchanges updates with that peer using either of the following methods,

1. Pull
2. Push
3. Hybrid

In pull method, a peer selects another peer randomly that lacks the information to be exchanged. Then the trigger is disseminated from the first peer to the lacking peer. It the process is reversed. That is the lacking peer asks for the information from the holding peer, then it is called push method. If the information diffusion is in both ways, then it is called hybrid method[5]. There are many studies that have shown the effectiveness of the entropy-oriented benefit like in [7-18] where authors have shown that different approaches can lead to better performance results. In [19-29] authors have shown that community-oriented neighbouring feedback can provide better and optimized performance response of delay-bounded energy-aware bandwidth allocation scheme in wireless networks. Authors have shown that by exploiting movement synchronization to increase end-to-end file sharing efficiency for delay sensitive streams in vehicular P2P devices using epidemic diffusion can drastically increase the performance of the system. Authors in [30-33] have investigated the efficient parallel data processing in the cloud using also an epidemic formality whereas in [34] authors have targeted -using diffusion and caching- at the same time energy consumption optimization through pre-scheduled opportunistic offloading in wireless devices.

## 5.2 SPATIAL-POPULARITY BASED INFORMATION DIFFUSION OR SPID

A peer to peer information diffusion scheme used in ad hoc mobile network is SPID or Spatial-Popularity based Information Diffusion model. There are constraints on bandwidth and mobility in mobile ad hoc networks. Sharing whole content within a single transmission is not always possible as the mobile nodes moves among different locations. A solution to this problem has been developed by SPID or Spatial-Popularity based Information Diffusion. The solution allows segmentation of the content into smaller pieces and exchanges those pieces. In case of mobile ad hoc networks, a node cannot predict the content from every other nodes. The SPID scheme helps in determining the urgency in dissemination of content based on the neighborhoods and necessity. The substantial segmentation policy and popularity of the dissemination in case of information exchange reduces the acquiring time related to content. There are extensive studies and simulations on this topic [6].

## 5.3 P2P SOCIAL NETWORK OVERLAY

As, it has been already told, social networks play a crucial role in information diffusion. OSN or Online Social Networks provide new ways of designing and presenting methods for information retrieval and diffusion. There are efficient topologies that helps in matching the OSN graph and P2P based topology model. The diameter of the graph is reduced by the topology and the route length is estimated by power law. There is diverse range of P2P oriented OSN systems that can fit into the application of this protocol. The basis of the topology is overlay. Search techniques in the model helps in efficient information diffusion and searching of data. The topology also helps in reducing the cost for signaling. [7]

# [6]. ONLINE SOCIAL NETWORKS AND INFORMATION DIFFUSION

OSN or Online Social Networks have millions of users from Internet from different corners of the globe. They have access to wealth of information from Internet. They can access these information real faster. In case of information diffusion in OSNs, the speed of information exchange based on different viewpoints are important considerations. OSNs provide information about events, incidents, interests and many more.

However, from the computer science viewpoints, following aspects are more important,

- The most popular topics of information that are discussed over the OSNs.
- Details of the information diffusion that may happen in future.
- Members of an OSN network that plays important roles in information diffusion.

The goals of this process is reviewing the developments of the OSN and information diffusion that can simplify the views related to a topic.

There will be dedicated web servers behind an OSN. An OSN or Social networking site has users. They can create profiles and exchanges messages. The users are explicitly connected with each other through 'social relationships'. The de factor of an OSN is, it can be thought as a content system with user generated contents. Where users can share information and communicate with each other.

An OSN is represented by a graph. The nodes of the graph are the users of the OSN and the edges are the relationships. The edges can be either directed or not, based on the relationship management of the OSN. For example, Facebook has bidirectional edges called 'friendship' on the other hand 'Twitter' uses unidirectional 'following' relationship. Messages from users are published and shared. A typical message will have an author, some test, timestamp and may be references to some other users. [8]

overlay in P2P social networks etc. Finally there is a discussion on online social media and information diffusion in general.

# [7]. CONCLUSION

In this report, there is detailed discussion on information diffusion and many related topics related to it. In the introductory section, it outlines the next parts of the report. Then there is brief discussion on information discussion from general and technical viewpoints. It is followed by a discussion on different kind of networks and their relations with information diffusion. The report is mainly focused on peer to peer models and information diffusion. So, it describes the models like epidemic model, self-organizing models etc. in peer to peer information diffusion. There are discussions on anti-entropy, SPID,

# [8]. REFERENCES